# Analysis of Carrier Frequency Selective Offset Estimation - Using Zero-IF and ZCZ In MC-DS-CDMA

Venkatachalam. Karthikeyan
Department of ECE, SVSCE, Coimbatore, India
karthick77keyan@gmail.com

Jeganathan. Vijayalakshmi.V
Department of EEE, SKCET, Coimbatore, India
vijik810@gmail.com

*Abstract*—A new method for frequency synchronization based upon Zero-Intermediate Frequency (Zero-IF) receiver and characteristics of the received signal's power spectrum for MC-DSCDMA-Uplink system is proposed in this paper. In addition to this, employing Zero Correlation Zone (ZCZ) sequences, designed specifically for quasi-synchronous uplink transmissions, is proposed to exploit frequency and temporal diversity in frequency-selective block-fading channels. The variance for Carrier Frequency Offset (CFO) estimators of MC-DS-CDMA Uplink is compared with that of an OFDM system to estimate the CFO. Our study and results show that the MC-DS-CDMA system is outperforming the OFDM method.

## I. Introduction

Technological advances over the past two decades have led to the rapid evolution of the telecommunications industry. No longer limited to narrow-band voice signals, modern communications integrate voice, images, data, and video on a level that was once considered to be impossible. Basically, in cellular systems there are four main methods of multiple access schemes such as FDMA, TDMA, CDMA, and OFDM. The main purpose of multiple access schemes is used to achieve the following number of goals such as – to handle several numbers of users in the same channel without any mutual interference problem in it. Also, to maximize the range of the spectral efficiency and in addition to this its robustness, that is enabling the ease of handover/handoff between the cells. This OFDMA is based around OFDM which is a form of transmission use large number of close spaced carriers that are modulated with low rate data. Though the signal is orthogonal to one another there will absence of mutual interference. The major advantage of OFDM signal is to avoid the effect of reflection and ISI over noisy channel. The carrier spacing is equal to the reciprocal of the symbol period. To study and perform comparative analysis of carrier frequency offset estimation of two types of multiple access signals such as OFDM and MC-DS-CDMA. Furthermore, the proposed method is simpler to implement than the former methods for the reason that the correlation with the CP, the identical parts such as pilot carriers and the null carriers are not used.

## II. ANALYSIS OF CARRIER FREQUENCY OFFSET

The MC-DS-CDMA Ref. [14] is multiple access techniques for future networks physical layer due to its ability to support multiple access capability, robustness to frequency selective channels, high spectral efficiency and narrow band interference rejection. Our contribution is to suggest a coarse offset estimation scheme for MC-DS-CDMA system based on Zero-IF and on ZCZ sequence spreading codes Ref. [13]

### B. MC-CDMA

Multi-Carrier Code Division Multiple Access (*MC-CDMA*) is a multiple access scheme used in OFDM-based telecommunication systems, allowing the system to support multiple users at the same time. In addition to this the table.1 clearly depicts the comparative analysis of various parameters like spreading, spectral efficiency, PAPR and application as shown in Table I.

## C. MC-DS-CDMA

One way of interpreting MC-CDMA is to regard it as a direct-sequence CDMA signal (DS-CDMA) which is transmitted after it has been fed through an inverse FFT (Fast Fourier Transform) Ref.[11] Multi-Carrier Code Division Multiple Access (MC DS-CDMA) is a multiple access scheme used in OFDM-based telecommunication systems, allowing the system to support multiple users at the same time. In the MC-DS-CDMA technique, the serial-to-parallel converted data stream is multiplied with the spreading sequence, and then the chips belonging to the same symbol modulate the same subcarrier. Here the spreading is done in the time domain. Thus this MC-DS-CDMA technique plays vital role in uplink transmission of wireless communication than any other access. The Transmitter section of this MC-DS-CDMA is given by the generalized block diagram shown in Figure.1 In TF-domain spreading MC-DS-CDMA, assume that there are J ZCZ time-domain spreading sequences available of length Q chips, i.e., $a_j(t)$ ( $j = 0,....., J-1$). Each time-domain spreading sequence is associated with P frequency-domain spreading sequences of length $M_B S$, i.e., $c_{j,p}$ ( $j = 0,...., P-1$),
where S and $M_B$ are the number of frequency tones and fading blocks over which frequency-domain spreading is done, respectively Ref. [10] Each user is assigned one time-domain spreading sequence $a_j(t)$ and one frequency-domain sequence $c_{j,p}$. The maximum number of users supported by TF-domain spreading MC-DS-CDMA is given by N = JP.

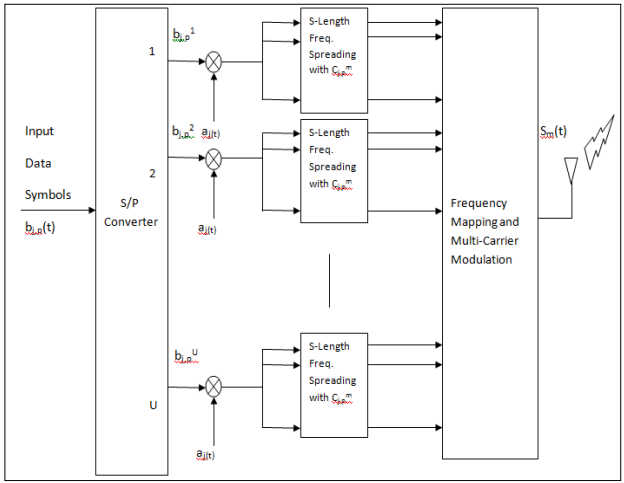

Figure 1. TF-domain spreading for MC-DS-CDMA

Fig.1 shows the transmitter schematic for a single-user in each fading block m. At the transmitter side, a symbol stream $b_{j,p}(t)$ with symbol period $T_{sym}$ is serial to parallel converted into U parallel branches in each fading block m. The above assumption implies that the coherence time $(\Delta t)_c$ of the channel is higher than but close to the symbol duration of $UT_{sym}$ (i.e. $(\Delta t)_c \geq UT_{sym}$). In this paper, we refer to the symbol duration of $UT_{sym}$ as a fading block. Note that, in practice, if the fading blocks are subject to correlated fading then interleaving may be employed over fading blocks at the transmitter, in order to guarantee the independent fading of the sub-carrier signals in each fading block. [10]Following time-domain spreading, the spread signal in each branch u is repeated on S parallel branches in each of MB fading blocks, where each parallel branch is multiplied by the corresponding chip value of a user-specific frequency-domain spreading sequence $c_{j,p} = [c_{j,p}[1], c_{j,p}[2], ...., c_{j,p}[SM_B]]^\tau$. In this analysis it is assumed that the same set of frequency-domain spreading sequences are used for each branch u. The U.S X 1 transmission vector after time-domain and frequency-domain spreading for each user, in fading block m can be expressed in (1)

$$Z_{j,p,m} = \begin{cases} c_{j,p}^m a_j(t) b_{j,p,1} \\ c_{j,p}^m a_j(t) b_{j,p,2} \\ . \\ . \\ c_{j,p}^m a_j(t) b_{j,p,U} \end{cases} \quad m=1,........,M_B \quad (1)$$

Where $c_{j,p}^m = [c_{j,p}[(m-1)S + 1], c_{j,p}[(m-1)S + 2],....., c_{j,p}[(m-1)S + S]]^\tau$, $m = 1,..., M_B$.

The cyclic prefix insertion matrix is expressed as
$T_{CP} = [I^\tau_{CP} I^\tau_{U.S}]^\tau$, where CP is the cyclic prefix length. The cyclic prefix removal matrix at the receiver can be expressed as $R_{CP} = [0_{U.SXCP} I_{U.S}]$. The matrix γ is defined as a U.S X U.S permutation matrix that maximally separates the data of each branch u in the frequency-domain, i.e., the $(x,y)_{th}$ element of is equal to 1 if $x = (y-1)\mod(U).S + [(y-1)/U]+1$ and equal to 0 otherwise. [9]The transmitter output for each user in each fading block m, from Fig.1, the transmitted signal can be expressed in (2)

$$s_m(t) = T_{CP} F^H_{U.S} \gamma z_{j,p,m}, m = 1,.....,M_B. \quad (2)$$

Where the above equation (2) represents the diagonal matrix of the channel frequency response in fading block m.

## D. Zero-IF Receiver

The zero-IF receiver, also known as a homodyne, synchrodyne or direct conversion receiver, is a special case of the super heterodyne receiver that uses an LO with the same frequency as the carrier. In Ref. [13] order for the detector to differentiate between signal components both above and below the LO frequency, zero IF receivers generate both In-Phase and Quadrature (IQ) signals. The following Fig. 2 which shows generalized block diagram of ZERO-IF receiver architecture. Further the main advantages of preferring this receiver is nothing but, Lower complexity and power consumption (no IF amplifier, no IF band-pass filter, or no IF local oscillator), which has the potential to reach the 'one chip goal' and No image frequency.

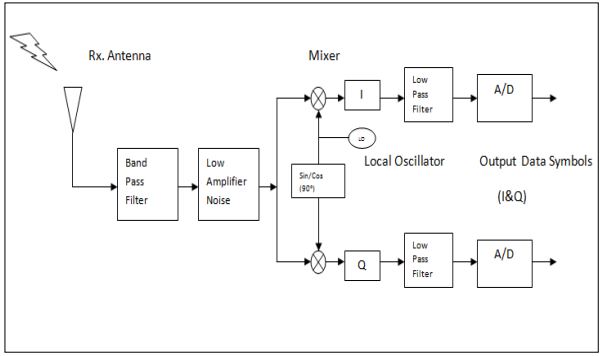

Figure 2. Zero-IF Receiver Architecture

*E. ZCZ Sequences*

Traditional orthogonal spreading sequences, such as orthogonal Gold sequences and Walsh-Hadamard sequences, exhibit non-zero off-peak cross-correlations, which limits the achievable performance in asynchronous or quasi-synchronous scenarios. However, ZCZ sequences exhibit an interference free window over which both the cross-correlation and autocorrelation function are zero Ref. [13]. Consequently, ZCZ sequences are able to suppress MUI in the quasi-synchronous uplink channel. We denote a set of ZCZ sequences as ($L_{seq}$, $M_{seq}$, $Z_0$) - ZCZ, where $L_{seq}$ is the sequence length, $M_{seq}$ is the sequence family size and $Z_0$ is the one-sided ZCZ length in chips. For a sequence set $\{s_q(q=0)\}^{Mseq}$ with family size $M_{seq}$, and sequence length $L_{seq}$, the periodic correlation characteristics of ZCZ sequences are defined in (3) and (4)

$$\Psi_{s,r}(\tau) = \sum_{l=0}^{L_{seq}-1} s_q[l] s_r[(l+\tau) \bmod (L_{seq})]$$

$$= \begin{cases} \eta L_{seq}, & \tau = 0, r = s \\ 0, & \tau = 0, r \neq s \\ 0, & 0 \leq |\tau| \leq \tau_0 \end{cases} \quad (3)$$

And

$$H = \frac{1}{L_{seq}} \sum_{l=0}^{L_{seq}-1} |s_r[l]| \leq 1 \quad (4)$$

Here each sequence element $s_q[l]$ is a complex number and $Z_0$ is the ZCZ length in chips. The parameter $\eta \leq 1$, is only equal to one when every sequence element $s_q[l]$ has unit amplitude, otherwise $\eta < 1$.. For a given ZCZ length, the theoretical upper bound is given in (5)

$$Z_0 \leq \frac{L_{seq}}{M_{seq}} - 1 \quad (5)$$

The above expression suggests that to obtain a large ZCZ length, the sequence length $L_{seq}$ needs to be considerably larger than the family size $M_{seq}$. Thus, the desirable properties of these sequences come at the cost of supporting only a small number of users, compared to other orthogonal sequences, such as Walsh-Hadamard sequences Ref. [9] we will focus on the power spectrum of the signal $X_{lb}(k)$ to estimating the blind carrier frequency offset without any identical parts and without addition of any redundant data. The power spectrum, $\rho^l(k)$, is given in (6)

$$\rho^l(k) = E[X_l(k) X_l^*(k)] \quad (6)$$

Where $X_l^*(k)$ is the complex conjugate of $X_l(k)$.
In the MC-DS-CDMA, the modulation of each symbol is constant over $N_f = N + N_{CP}$ samples, and $e^{j(\varphi i(n)-\varphi i(m))}$ is a stationary process. Therefore, its expected value is calculated as given in (7)

$$E[e^{j(\varphi_i(n)-\varphi_i(m))}] = \begin{cases} \frac{N_f - |n-m|}{N_f}, & |n-m| \leq N_f \\ 0, & \text{otherwise} \end{cases} \quad (7)$$

## III. RESULTS AND DISCUSSIONS

Figure 3 compares the variance theory of the CFO estimator with simulations for MC-DS-CDMA and OFDM systems for a symbol of 256 samples, according to the length of the cyclic prefix $N_{CP}$. Specifically, these components are pattern-dependent self-noise and AWG channel noise. Figure 4 and Figure 5 compare the simulated and theoretical variances as a function of SNR. At high SNRs, there is again a difference of approximately 10% between the simulated and the theoretical variances for the majority of the simulations. At very low signal to noise ratios, this difference increases until the simulated variance is approximately 70% larger than the theoretical variance at an SNR. As the SNR increases, less improvement in variance performance is observed in Figure 4 and Figure 5 which is consistent with the mathematical Variance expression as the contribution from the AWG component of the noise becomes overshadowed by the estimator self-noise. Figure 4 shows the effects of number of symbols used to estimate the CFO at an SNR of 10dB. At all $N_{sym}$ values, the variance of the CFO in MC-DS-CDMA system is roughly 70% smaller than the OFDM variance and roughly 10% larger than the theoretical variance. Fig 5 shows the effects of the AWGN component in variance of the CFO as a function of SNR. At very low SNR (< 10 dB), the variance is approximately larger than the theoretical variance, but is smaller than the OFDM variance, while for SNRs above 10dB, the theoretical variance predicts the observed simulated. Figure 5 shows the performance results using the optimal ML detector and the MMSE block linear detector at the receiver. One observes from Figure 6 that there is an approximate 2.7 dB performance improvement for the optimal ML detector at a BER of $10^{-4}$ to $10^{-5}$.

TABLE .1 COMPARISONS OF MC-CDMA AND MC-DS-CDMA

| Spreading | Frequency | Time Direction |
|---|---|---|
| Spectral Efficiency | High | Low when multi-carrier modulation other than OFDM is used |
| PAPR | High in Uplink | Low in Uplink (Appropriate for uplink in multi-user system) |
| Application | Synchronous Uplink and Downlink | Asynchronous uplink and downlink |

## IV. SIMULATION RESULTS

Simulations are carried out to investigate the performance of the system by choosing BER as a figure of merit. In the simulations, the channel coefficients are constant during one symbol block duration (i.e. $UT_{sym}$), but change from one block to another one. One observes only a slight performance improvement, when the MMSE detector is employed at the receiver. Thus, the proposed system outperforms the system proposed in literature. Owing to the additional temporal diversity exploited in block fading conditions.

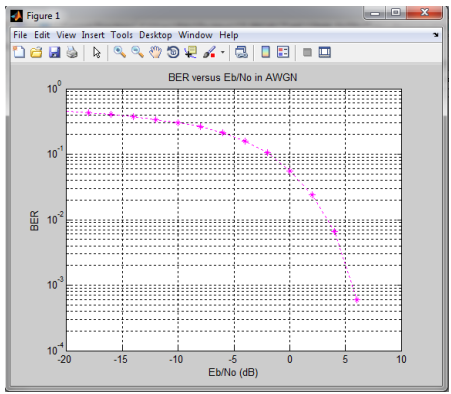

Figure 3. $N_{CP}$ for a symbol of 256 samples

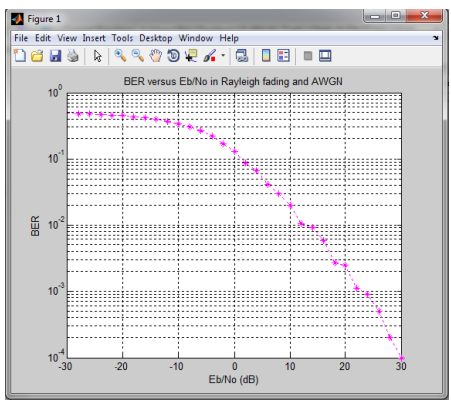

Figure 4. Effects of Number of symbols

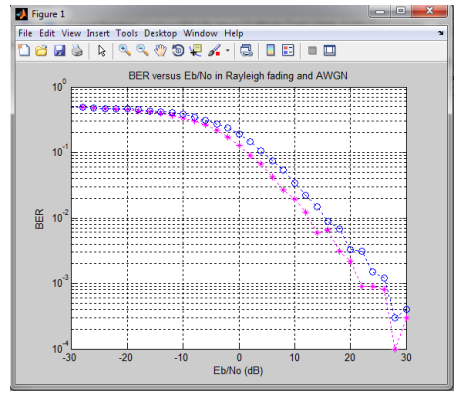

Figure 5. Effects of AWGN

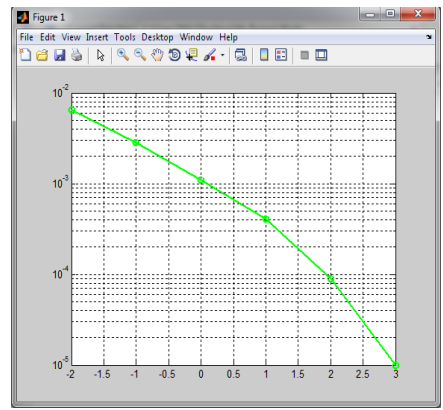

Figure 6. Performance of BER Vs Number of symbols

## V. CONCLUSION

Thus the objective of this project work was to develop and to analyze an algorithm for blind CFO recovery suitable for use with a practical zero-IF OFDM telecommunications system [14]. MC-DS-CDMA is more sensitive to carrier frequency offsets than other modulation techniques like QAM. CFOs significantly degrade the SNR at the output of the receiver.

**Venkatachalam. Karthikeyan** has received his Bachelor's Degree in Electronics and Communication Engineering from PGP college of Engineering and technology in 2003 Namakkal, India. He received Masters Degree in Applied Electronics from KSR college of Technology, Erode in 2006. He is currently working as Assistant Professor in SVS College of Engineering and Technology, Coimbatore. She has about 7 years of teaching Experience.

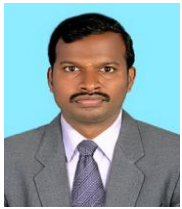

**Jeganathan. Vijayalakshmi** has completed her Bachelor's Degree in Electrical & Electronics Engineering from Sri Ramakrishna Engineering College, Coimbatore, India She finished her Masters Degree in Power Systems Engineering from Anna University of Technology, Coimbatore. She is currently working as Assistant Professor in Sri Krishna college of Engineering and Technology, Coimbatore. She has about 5 years of teaching Experience.

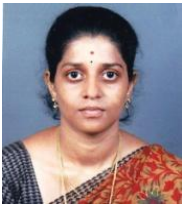